\documentclass[aps,prstab,fleqn,preprint,nofootinbib,superscriptaddress,groupedaddress]{revtex4-1}

\usepackage{bm}
\usepackage{graphicx}
\usepackage{adjustbox}
\usepackage{amsmath}
\usepackage{MnSymbol}
\usepackage[section]{placeins}
\usepackage{color}\definecolor{gray}{gray}{0.5}

\begin{document}
\title{Spin orbit torque nano-oscillators by dipole field-localized spin wave modes
}
\author{Chi Zhang$^\ast$}
\author{Inhee Lee}
\author{Yong Pu$^\dagger$}
\author{Sergei A. Manuilov}
\author{Denis V. Pelekhov}
\author{P. Chris Hammel$^\ast$}
\affiliation{Department of Physics, The Ohio State University, Columbus, OH 43210, USA}
	
\date{\today}
\begin{abstract}
We demonstrate a high-quality spin orbit torque nano-oscillator comprised of spin wave modes confined by the magnetic field by the strongly inhomogeneous dipole field of a nearby micromagnet. This approach enables variable spatial confinement and systematic tuning of magnon spectrum and spectral separations for studying the impact of multi-mode interactions on auto-oscillations. We find these dipole field-localized spin wave modes exhibit good characteristic properties as auto-oscillators---narrow linewidth and large amplitude---while persisting up to room temperature. We find that the linewidth of the lowest-lying localized mode is approximately proportional to temperature in good agreement with theoretical analysis of the impact of thermal fluctuations. This demonstration of a clean oscillator with tunable properties provides a powerful tool for understanding the fundamental limitations and linewidth contributions to improve future spin-Hall oscillators.
\end{abstract}
\pacs{}
\def\thefootnote{$\ast$}\footnotetext{hammel@physics.osu.edu, zhang.1390@osu.edu}
\def\thefootnote{\dagger}\footnotetext{Current address: Nanjing University of Posts and Telecommunications, Nanjing 210023, China}
\maketitle
%
\section{Introduction} 
Spin Hall effect (SHE)-spin torque can drive magnetic auto-oscillations or excite propagating spin waves \cite{PSW1,PSW2,ReviewPSW,NonlinearDamp2,R1I2} in a ferromagnet \cite{demidov_magnetic_2012,liu_magnetic_2012,collet_generation_2016}, potentially enabling a dc current-tunable microwave source \cite{kiselev_microwave_2003,ReviewMW}, neuromorphic computing \cite{ReviewNeuromorphic,MinSyn} and spin orbit torque magnonics \cite{PSW1,PSW2,ReviewPSW,NonlinearDamp2,R1I2}. However, nonlinear magnon scattering can degrade the quality of spin-Hall oscillators broadening linewidth, and hampering the achievement of auto-oscillation \cite{collet_generation_2016,duan_nanowire_2014,IgorGiant}. A ferromagnetic film with spatially extended dimensions harbors a large number of degenerate spin wave modes which can enable nonlinear magnon scattering that redistributes energy between modes, hampering coherent oscillation of a desired mode \cite{duan_nanowire_2014,demidov_control_2011,DemidovReview}. These adverse effects can be reduced either by avoiding the degeneracy \cite{DemidovOOP,demidov_magnetic_2012} \textit{e.g.} via spatial localization or by suppressing the nonlinear mode coupling \cite{NonlinearDamp1,NonlinearDamp2}. Spatial localization in, \textit{e.g.}, nanoconstrictions and planar nano-gap contacts \cite{demidov_magnetic_2012,demidov_nanoconstriction-based_2014,demidov_spin-current_2015,liu_spectral_2013,demidov_spectral_2015,Akerman_SingleLayer,Akerman_LongRangeSychonize,DW,Bipolar,TRMOKE,CMOS,Origin} produces discrete modes whose frequencies lie below the spin wave dispersion for the extended sample thus reducing scattering channels \cite{demidov_magnetic_2012,duan_nanowire_2014,R1I2} and enabling auto-oscillation. Alternative to geometrical confinement, a localized region of reduced internal magnetic field generated by the dipole field of a nearby micromagnetic particle confines the spin wave modes in a nanoscale magnetic field well \cite{adur_damping_2014,chia_nanoscale_2012,lee_nanoscale_2010,OscillatorChiZhangEngineering} thus enabling auto-oscillation. Dipole-field localized modes offer advantages as spin-Hall oscillators \cite{OscillatorChiZhangEngineering}: First, in contrast to most geometrically confined oscillator modes, our field  confinement offers in-situ tunability. In particular, the lateral confinement and field well depth are tunable by particle parameters---size, moment, particle-sample separation---the latter of which can be altered in-operando, and is continuously controllable if the particle is mounted on a cantilever whose height is variable. This provides an in-situ tunability of the magnon spectrum and mode separations, and hence tuning of nonlinear magnon scattering to control performance of the oscillators. Second, these dipole-field localized modes are confined by magnetic field rather than by sample boundaries thus avoiding artifacts from fabrication imperfection, so they can provide cleaner confinement.

Here, we demonstrate spin-Hall auto-oscillations of localized spin wave modes spatially confined by the magnetic dipole field of a nearby Ni particle above a Py/Pt stripe. Multiple well-resolved localized modes are driven into narrow-linewidth, large-amplitude auto-oscillations persistent up to room temperature (RT). The auto-oscillating localized modes exhibit a minimum linewidth of 5 MHz at 80 K which increases to 18 MHz at RT. We observe a linewidth for the first localized mode that is approximately proportional to temperature and primarily limited by thermal fluctuations. This high quality oscillator with its in-operando tunability of spatial confinement provides a powerful tool for systematic study of the fundamental aspects of spin-Hall oscillators, including the impact of multi-mode interactions, for advancing the future oscillators.
\section{METHODS}
\begin{figure}[!t]
	\includegraphics[width=0.65\columnwidth]{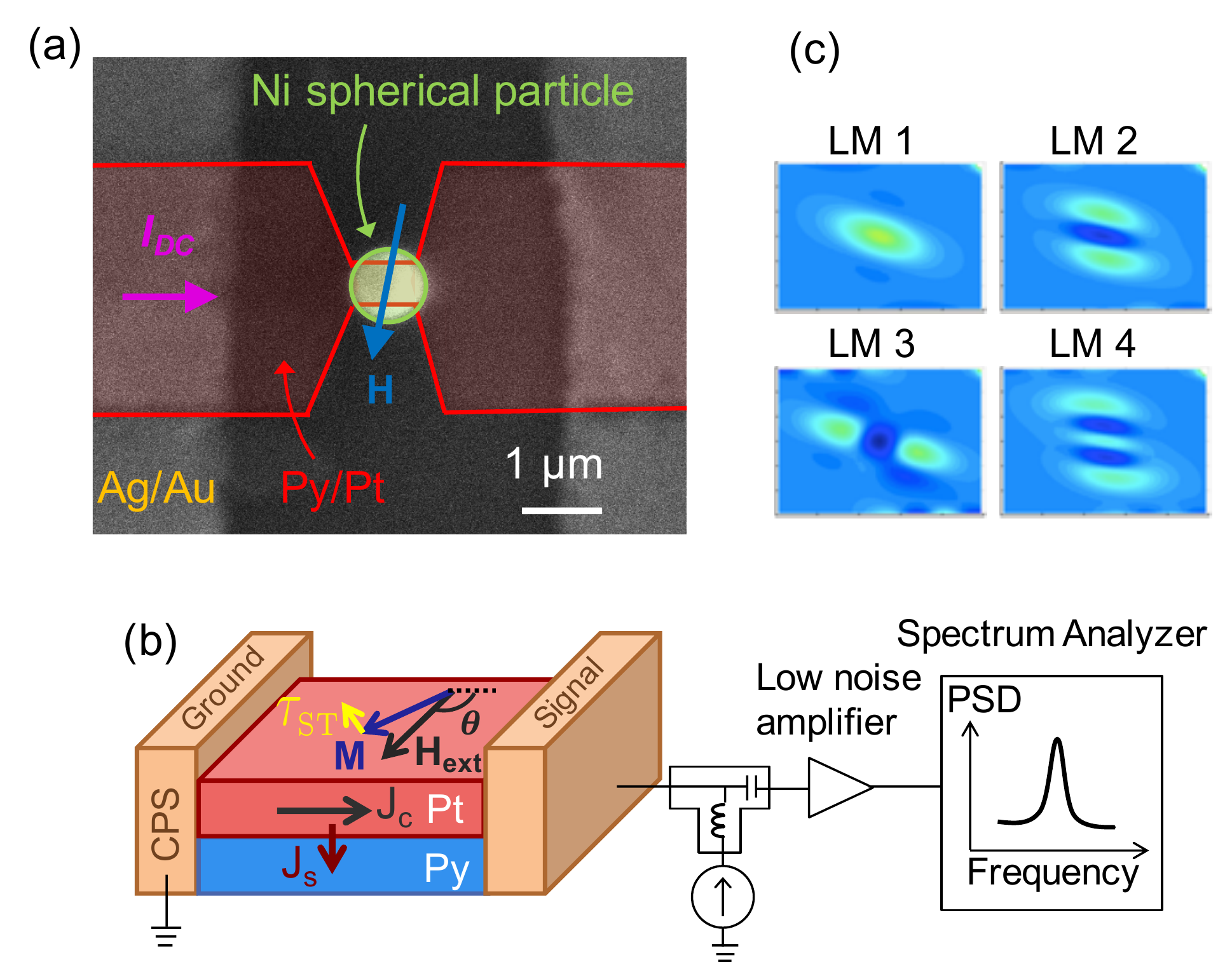}
	\caption{\label{Figure1} (a) Scanning electron micrograph of the Py/Pt strip device (outlined in red) with a Ni spherical particle (outlined in green) glued on top. The active region of the device is the center 700 nm $\times$ 1 $\mu\rm m$ region of the nanostrip underneath Ni particle. (b) Schematic of auto-oscillation measurement setup. The external field $H_{\text{ext}}$ is applied in the plane, at an angle $105^{\circ}$ with respect to the direction of the current flow along the length of the strip. The magnetization $M$ and spin torque $\tau_{\text{ST}}$ are drawn for the bottom Py layer. (c) Spatial mode profiles of transverse magnetization of the first four lowest-lying localized modes in the 700 nm $\times$ 1 $\mu\rm m$ active region, labeled as LM1, LM2, LM3 and LM4.}
\end{figure}%
Py (5 nm)/ Pt (5 nm) films are deposited on an undoped-Si substrate by e-beam evaporation. The Py/Pt strips are defined with e-beam lithography using Cr as a hard mask \cite{DuanThesis_2013} for ion milling. A Ti/Ag/Au coplanar stripline (CPS) is fabricated on the Py/Pt strip via photolithography. Figure 1(a) shows a scanning electron microscopy image of our device. 3 $\mu\rm m$-width strips provide tapered transitions to the 700 nm $\times$ 1 $\mu\rm m$ region of interest in the center where current density is much higher. This geometry reduces heating and contact resistance. A 150 nm-thick $\text{Al}_2\text{O}_3$ spacer deposited on the device defines its separation from a 1 $\mu\rm m$ Ni particle located on top, in the center of the active region, with $\pm$ 200 nm accuracy in the strip width direction [see supplement for further details]. We measured two nominally identical devices labeled as Device \#1 and Device \#2. They show similar results with minimal device-to-device variation [see supplement]. We report measurements performed at temperatures from 80 K to RT with an external field $H_{\text{ext}}$ applied in the film plane at an angle $105^{\circ}$ with respect to the orientation of the current flow along the length of the strip. As depicted in Figure 1(b), the spin Hall effect \cite{DYakonovM.I.;Perel1971,hirsch_spin_1999} in Pt converts a charge current density $J_C$ into a spin current density $J_S$, which exerts an anti-damping spin torque $\tau_{\text{ST}}$ \cite{berger_emission_1996,slonczewski_current-driven_1996} on the Py magnetization. We apply dc charge currents exceeding the critical current needed to fully compensate the damping and excite the auto-oscillation. The microwave power emitted by the device is detected via anisotropic magnetoresistance (AMR) effect, and measured by a low-noise amplifier and spectrum analyzer \cite{duan_nanowire_2014,KiranThadaniThesis_2009}.

We use a nearby micromagnetic particle to introduce a spatially localized minimum in the internal static field---a field well---in the sample \cite{adur_damping_2014,chia_nanoscale_2012,lee_nanoscale_2010,OscillatorChiZhangEngineering}. The lowest-frequency spin wave modes at the bottom of the well are discrete localized modes (LM) whose frequencies lie well below the spin wave continuum. In addition, a large number of spin wave modes, both sample-confined modes and highest-order localized modes with high mode densities, reside at the top of the field-well at much higher frequencies. The depth of the field-well, typically several hundreds of Gauss, determines the spectral separation between lowest-lying localized modes and the high density higher-frequency modes. The spatial confinement provided by the field-well determines the spectral separations between the various discrete localized modes. The four lowest-frequency localized modes are localized width modes LM1 ($n$ = 1, $m$ = 1), LM2 ($n$ = 2, $m$ = 1), a localized length mode LM3 ($n$ = 1, $m$ = 2), and a localized width mode LM4 ($n$ = 3, $m$ = 1) \cite{OscillatorChiZhangEngineering}. Figure 1(c) shows micromagnetic simulated mode profiles of the lowest-four localized modes in the 700 nm $\times$ 1 $\mu\rm m$ active region. The mode area of LM1 is about 350 nm $\times$ 600 nm.
\section{RESULTS and Discussion}
\subsection{Auto-oscillation of localized modes: characteristic properties}
\begin{figure}[!htb]
	\includegraphics[width=0.55\columnwidth]{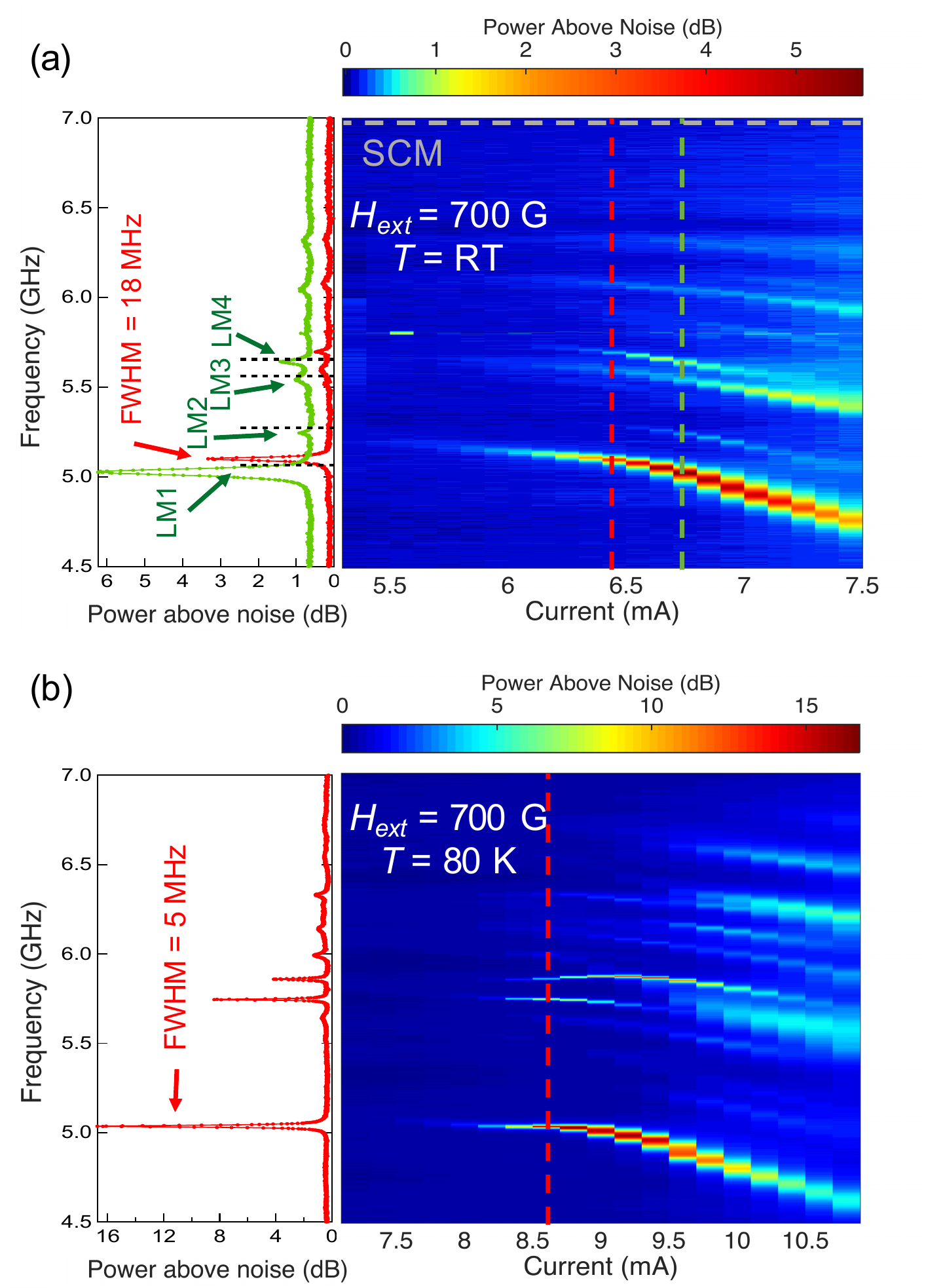}
	\caption{\label{Figure2} Auto-oscillation color maps (as a function of bias current) and representative spectra (corresponding to vertical dashed line on the color maps) of microwave emission generated by Py/Pt strip device with a Ni particle (Device \#2) at $H_{\text{ext}}$ = 700 G, (a) $T$ = RT and (b) $T$ = 80K. Plotted on the left are power above noise in units of dB with background subtraction from spectra obtained at 0 mA. In Figure 2(a), the power baseline of the spectrum (green) is manually offset. Black dashed lines in the spectra mark resonance frequencies at 700 G obtained from micromagnetic simulations, enabling identification of the localized modes LM1 to LM4. Dashed horizontal line (gray) in the color map in (a) indicates the frequency of sample-confined modes around 7 GHz extracted from ST-FMR at high current at RT.}
\end{figure}%
Figure 2(a) shows an auto-oscillation color map (as a function of bias current) and representative spectra of Device \#2 at $H_{\text{ext}}$ = 700 G, $T$ = RT. Below 5.4 mA, there is no signal. Above 5.4 mA, we see a group of several auto-oscillatory modes, corresponding to localized modes that are driven into auto-oscillations [see subsequent further analysis]. The measurements here are done at RT, demonstrating that localized modes function well as RT auto-oscillators which distinguishes them from some existing spin-Hall oscillators \cite{duan_nanowire_2014,liu_spectral_2013}. If we focus on the first, \textit{i.e.}, lowest-frequency, localized mode, the full width at half maximum (FWHM) linewidth, fitted on a linear power scale, decreases to a minimum value of 5 MHz at 8.5 mA at 80 K [Figure 2(b)], which is a typical value for good single spin-Hall oscillators \cite{duan_nanowire_2014,liu_spectral_2013,demidov_nanoconstriction-based_2014,demidov_spin-current_2015,demidov_spectral_2015,yang_reduction_2015,Akerman_20nm,Akerman_PyW,MinLine1,MinLine2,MinLine3,MinLine4}, and a much lower value can be achieved in mutually synchronized oscillator arrays \cite{MinSyn}. And even at RT, a linewidth of 18 MHz is still a small value, showing that oscillators derived from dipole field localized modes are highly coherent. The signal amplitude initially increases with current. The maximum power of the first localized mode is 0.27 pW at RT, and is 2.5 pW at 80 K [see Figure 3(c) or 4(a)], comparable to the 4.6 pW signal size of the bulk mode ($H_{\text{ext}}$ = 700 G, $T$ = 80 K) in a bare 700 nm $\times$ 1 $\mu\rm m$ Py/Pt strip device without a particle. This observation of pW signals from localized modes, even though the signal is only from the reduced local mode area and suffers from current shunting indicates our localized modes are driven into large-amplitude oscillations. The higher-order localized modes become more evident at lower temperatures. A few higher-order localized modes grow into large signals, and are as sharp as the first localized mode.
\subsection{Mode nature}
\begin{figure}[h]
	\includegraphics[width=0.7\columnwidth]{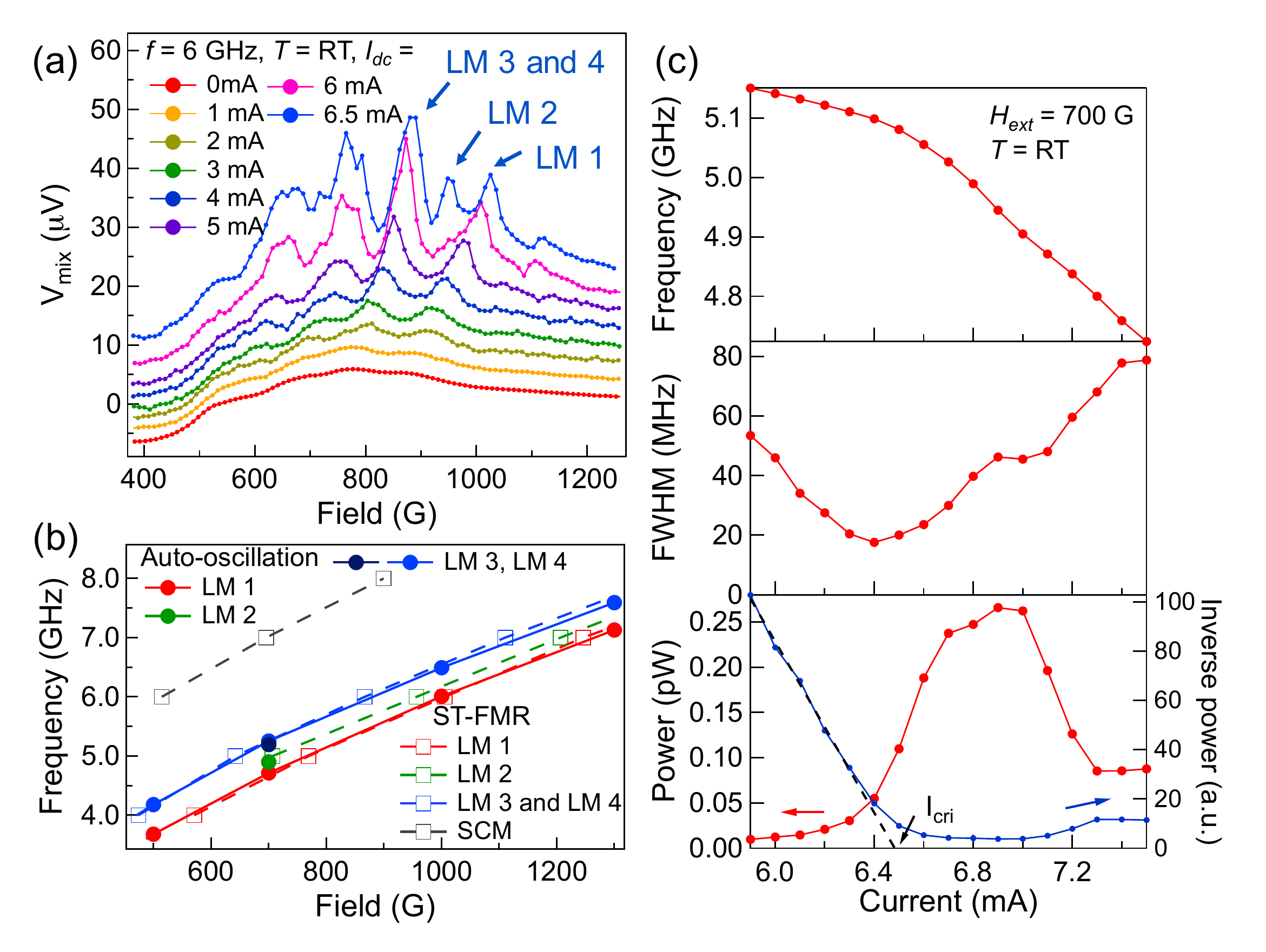}
	\caption{\label{Figure3} (a) ST-FMR spectra of Py/Pt strip device with a Ni particle (Device \#1) for several currents at RT and $f$ = 6 GHz. (b) Frequency versus field resonances of LM1 to LM4: (open squares) eigenmodes measured by ST-FMR and (filled circles) auto-oscillatory modes (Device \#1). The resonances of sample-confined modes from ST-FMR are shown as a reference. The resonances were extracted at 5.5 mA from ST-FMR and 6.5 mA from auto-oscillation for LM1, 3 and 4. The values for LM2, which appears only at higher currents, were extracted at 6.5 mA from ST-FMR and 6.8 mA from auto-oscillation. (The current values are selected to minimize the nonlinear frequency shift and consider the effect of RF current.) (c) Bias current dependence of the spectral frequency, FWHM linewidth, integrated power and inverse integrated power of LM1 auto-oscillation (Device \#2). The dashed line in Panel (c) represents a linear fit to inverse power used to identify the critical current \cite{InversePower}.}	
\end{figure}%
Here, we compare spin torque ferromagnetic resonance (ST-FMR) \cite{liu_spin-torque_2011}, auto-oscillation and micromagnetic simulation to determine mode nature. Figure 3(a) shows the ST-FMR spectra of Device \#1 at 6 GHz with increasing currents. At currents above 2 mA, we see several well-resolved localized modes in the 600--1000 G field range. The localized modes are not individually resolvable below 2 mA (including negative currents), because the linewidth of the modes, controlled by spin Hall torque, exceeds or is just comparable to their separation \cite{OscillatorChiZhangEngineering}. The signal around 500 G comes from sample-confined modes (SCM) that extend throughout the whole sample but are modified by the field of the particle \cite{OscillatorChiZhangEngineering}. Increasing current shifts the resonances of localized modes to higher fields due to Oersted field \cite{OscillatorChiZhangEngineering}, Joule heating \cite{OscillatorChiZhangEngineering} and more importantly nonlinear enhancement of the cone angle at high currents \cite{OscillatorChiZhangEngineering,slavin_nonlinear_2009}. We compare resonances of several of the lowest-lying localized modes from ST-FMR and auto-oscillation measurements [see supplement for auto-oscillation color maps of Device \#1] to confirm the nature of the auto-oscillatory modes. The frequency versus field dispersion from both measurements coincide with each other [Figure 3(b)], which confirms that the auto-oscillatory modes arise directly from the localized modes that are confined by the particle-field. Then, we perform micromagnetic modeling \cite{adur_damping_2014,Adur2015,du_experimental_2014,obukhov_local_2008,OscillatorChiZhangEngineering} at $H_{\text{ext}}$ = 700 G and compare the resonances with auto-oscillation peaks in Figure 2(a). The simulated resonance frequencies for LM1 to LM4 are shown as dashed lines in Figure 2(a), identifying the first four localized modes [see supplement for mode identification at other temperatures]. The parameters used in the simulation are: 4$\pi M_{\text{eff}}$ = 8.9 kG, determined from the in-situ Py/Pt film, the effective magnetization of the unsaturated Ni particle 4$\pi M_{\text{Ni}}$ = 2400 G which is a free parameter \cite{chia_nanoscale_2012,OscillatorChiZhangEngineering}, and the modeling is done at $I_{\text{dc}}$ = 0. The corresponding localized modes are also labeled in ST-FMR spectra [Figure 3(a)]. We note that the energies of LM3 and LM4 are close, so that their peaks are not distinguishable in ST-FMR. Mode LM2 appears at high currents in ST-FMR and does not present a strong auto-oscillation signal. This could be because the spatial $180^{\circ}$ phase reversal in the transverse magnetization of $n$=2 mode profile causes partial cancellation in some of the AMR signals \cite{AMRphase}. Further details about the observed localized modes in auto-oscillation spectra can be complicated. For LM1--4, we observe all four modes at 700 G, and only observe LM1 and one of the LM3,4 peaks at several other fields [see supplement for color maps at different fields].

We comment on the modes excited in auto-oscillations. Only the localized modes and second harmonics of the localized modes [see supplement for full spectrum] are observed. We observe no auto-oscillation signals at the frequencies corresponding to sample-confined modes, while there are signals around those resonance conditions in ST-FMR spectra. This is probably a consequence of the large degeneracies of the modes in this frequency range, both the sample-confined modes and highest-order localized modes.
\subsection{Evolution of properties with current}
To systematically study the evolution of the auto-oscillation with current, we subtract a background obtained from the spectrum at 0 mA on a linear scale. We then determine the characteristic parameters of LM1 at $H_{\text{ext}}$ = 700 G, $T$ = RT (Device \#2) and plot them as a function of current, as shown in Figure 3(c). The current-evolution trends that we observed are similar to ones in existing spin-Hall oscillators \cite{duan_nanowire_2014,liu_spectral_2013}. A critical current $I_{\text{cri}}$ is defined as that at which damping is fully compensated by the spin torque. We extract the critical current $I_{\text{cri}}$ via a linear fit of inverse power vs. current \cite{InversePower}. Microwave emission signals can be observed even before the $I_{\text{cri}}$, due to thermal fluctuations \cite{slavin_nonlinear_2009}. Above the critical current, the magnetization precession enters the nonlinear regime as its cone angle increases. In the nonlinear regime, the frequency $\omega$ is a function of the power $p$: $\omega \approx \omega_0 + Np$ \cite{slavin_nonlinear_2009}. Where, $\omega_0$ is the frequency in small-angle approximation. $N$ is the nonlinear frequency shift coefficient, and is typically negative for in-plane field geometries. With increasing current, the power increases leading the frequency to shift to lower values. The shift is slow below $I_{\text{cri}}$, because the power is small from thermally excited precessions, becoming rapid above $I_{\text{cri}}$ as a consequence of the much larger auto-oscillation power. Below $I_{\text{cri}}$, it is in linear regime, so the linewidth is linearly reduced by the antidamping torque. If we only consider nonlinear auto-oscillation theory of a single mode as in conventional spin torque oscillators \cite{slavin_nonlinear_2009}, the linewidth would be expected to remain small after the linear reduction. However, here at high currents, linewidth broadens, and power drops, which is experimentally observed in other spin-Hall oscillators \cite{duan_nanowire_2014,liu_spectral_2013}, and could be due to nonlinear magnon scattering through the remaining scattering channels \cite{duan_nanowire_2014}.
\subsection{Temperature dependence}
\begin{figure}[!t]
	\includegraphics[width=0.45\columnwidth]{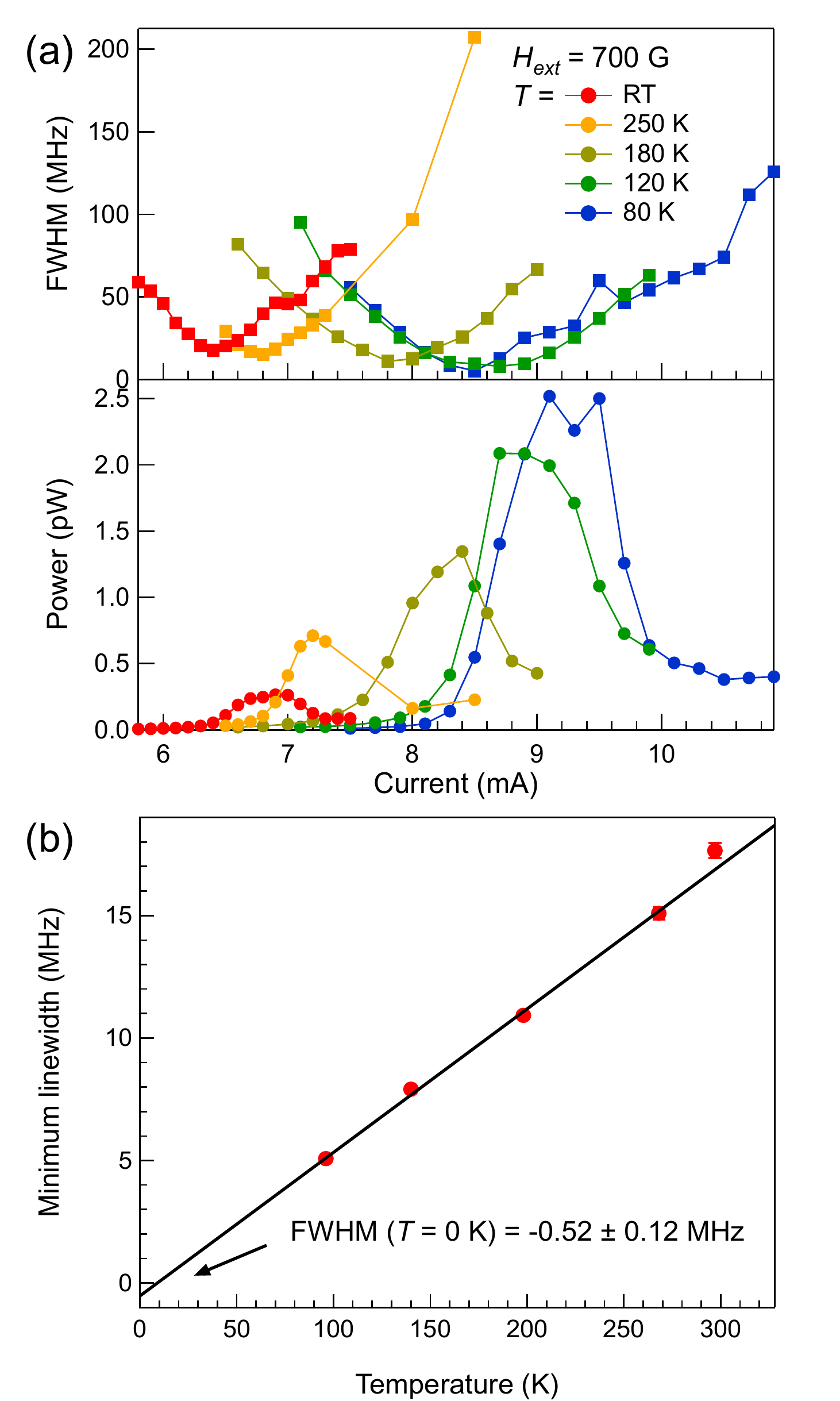}
	\caption{\label{Figure4} Linear temperature dependence of linewidth (Device \#2). (a) Bias current dependence of the FWHM linewidth and integrated power of LM1 auto-oscillation at different bath temperatures. (b) Actual temperature $T_a$ dependence of the minimum FWHM linewidth of LM1 auto-oscillation. The error bars in (b) are smaller than the data point.}	
\end{figure}%
We further measure current evolution of auto-oscillations at $H_{\text{ext}}$ = 700 G at different temperatures (Device \#2). The mode structures are qualitatively the same as the ones in Figure 2 [see supplement]. Figure 4(a) shows the extracted FWHM linewidth and integrated power of the first localized mode as a function of current for different temperatures. As temperature increases, the required bias current decreases, likely due to an increased spin current conversion efficiency by SHE \cite{liu_spectral_2013}. The maximum power decreases with increasing temperature, which is generally observed in other spin-Hall oscillators \cite{duan_nanowire_2014,liu_spectral_2013}.

To determine the temperature dependence of the oscillation linewidth, we exploit the calibrated dependence of the device's resistance on temperature. This gives the actual temperature $T_a$ of the device in the presence of inevitable Joule heating above the bath temperature $T_b$ by high drive currents which is obtained by measuring the resistance of device \cite{liu_spectral_2013} [see supplement]. We analyze the behaviors at current corresponding to the minimum linewidth \cite{Linear_RHLiu,PRB_2009_LinearWithIntercept} as a function of the actual temperature at $H_{\text{ext}}$ = 700 G. Figure 4(b) shows that the linewidth of the first localized mode varies linearly with temperature. The damping has been fully compensated, but the linewidth of auto-oscillators is a measure of the coherence of the system and the precession periodicity \cite{JackSankeyThesis_2007,KiranThadaniThesis_2009}. At finite temperatures, thermal fluctuations deflect the magnetization transverse to (amplitude fluctuations) and along (phase fluctuations) its precession trajectory, and leads to a spread of the frequencies \cite{slavin_nonlinear_2009,AkermanReview_IEEE}. The generated linewidth due to thermal fluctuations is expected to be proportional to the temperature \cite{slavin_nonlinear_2009}. Such a theory accounts well for our data, and indicates that the linewidth of our localized mode oscillator is limited dominantly by thermal fluctuations. A linear fit to the linewidth data yields an intercept of -0.52 $\pm$ 0.12 MHz, close to zero. We compare the temperature-dependence of our observed linewidth with existing auto-oscillators. Ours is the only report of a linear temperature-dependence of linewidth among spin-Hall based oscillators \cite{duan_nanowire_2014,liu_spectral_2013,demidov_nanoconstriction-based_2014,demidov_spin-current_2015,yang_reduction_2015,demidov_spectral_2015,Akerman_20nm,Akerman_PyW} with one exception \cite{Linear_RHLiu}. Even among conventional spin torque oscillators, most temperature dependence studies present unexpected behaviors, more complex than proportional, indicating the existence of other broadening effects \cite{PRB_2012_Nonlinear,PRB_2005_Exponential,PRB_2009_DependLittle,APL_2006_DependLittle}. Existing reports \cite{PRB_2009_LinearWithIntercept,APL_2012_LinearWithSaturation,APL_2012_LinearWithIntercept} of linear temperature dependencies exhibit large negative or positive intercept. The non-zero intercept is attributed to an additional temperature-independent broadening mechanism \cite{PRB_2009_LinearWithIntercept,APL_2012_LinearWithSaturation,APL_2012_LinearWithIntercept}, possibly due to inhomogeneities. Our intercept (both its magnitude and as a percentage of total linewidth) is the smallest among all these results which may reflect the avoided fabrication-induced broadening afforded by magnetic confinement. We note that the influence of edge damage can be studied by moving the field well toward the edge of the sample to observe the impact of the edge on oscillator performance. Our method of scannable magnetic confinement provides a new approach to understanding problems like these and examining the hypotheses. Additionally, our linewidth, approximately proportional to temperature, agrees well with Slavin's thermal fluctuations theory. These results suggest our dipole-field confined oscillator closely approaches a nearly ideal oscillator.

Beyond fabrication edge imperfections, thermally activated mode hopping \cite{liu_spectral_2013} (causing exponential temperature dependence) and nonlinear magnon scattering \cite{demidov_control_2011,duan_nanowire_2014} can contribute to the linewidth; field-localization of modes is advantageous in relation to these contributions as well. First, localized modes confined by the particle-field have large spectral separation from the high density of modes at higher frequencies as well as increased spectral separation between different localized modes, and so increase the energy barrier for mode hopping. Second, the tight spatial confinement increases the separation between the discrete lowest-lying localized modes thus weakening the mechanism of nonlinear magnon scattering.
\section{CONCLUSIONS} 
In summary, a new type of spin-Hall oscillator that is spatially confined by a localizing magnetic field is demonstrated. We observe multiple dipole field-localized spin wave modes driven into auto-oscillation by spin-Hall torque. These localized modes oscillators exhibit good properties: narrow-linewidth, large-amplitude and persistence up to RT. The auto-oscillating localized modes exhibit a linewidth of 5 MHz at 80 K which increases to 18 MHz at RT. We observe that the linewidth of the first localized mode is approximately proportional to temperature, limited dominantly by thermal fluctuations. These results indicate a clean oscillator system which provides opportunity to study auto-oscillators in their intrinsic regime. The concurrence of good properties, proportionality to temperature and field-localization demonstrates the advantages of field-localization method.

For future applications, the dipole-field localization provides an alternative confining mechanism that can assist further studies and help resolve issues in spin-Hall auto-oscillators, \textit{e.g.} their spatial self-broadening \cite{OscillatorYIGBLS}. In addition, the demonstrated unique strength of localized modes spin-Hall oscillator here can be expanded by putting the particle on a cantilever to scan the localized mode oscillator ($x$ and $y$ directions) and tune well spatial confinement and depth ($z$-direction). We observed signals with sufficiently strong signal-noise-ratio that allows the strip to be widened to a few microns for imaging. Spatial confinement studies tuned by varying cantilever tip-sample separation can reveal the impact of multi-mode interactions on spin-orbit torque auto-oscillation. Finally, the particle-confined localized modes oscillator combined with spatial-scannability can be utilized to study synchronization \cite{Akerman_LongRangeSychonize} and interactions between spatially separate auto-oscillators by varying their lateral separation.
\begin{acknowledgments}
This research was primarily supported by the Center for Emergent Materials, an NSF MRSEC, under award number DMR-2011876. This work was supported in part by an allocation of computing time from the Ohio Supercomputer Center. We also acknowledge technical support and assistance provided by the NanoSystems Laboratory at the Ohio State University partially supported by DMR-2011876.
\end{acknowledgments}

\newpage
\renewcommand{\figurename}{Supplementary Figure}
\setcounter{figure}{0}
\begin{center}
	{\large \textbf{Supplementary Information\\Spin orbit torque nano-oscillators by dipole field-localized spin wave modes }}
\end{center}
\section*{A. Comparison between two devices with little device-to-device variation}
Device \#1 and Device \#2 are patterned on the same substrate with the same dimensions. The particles are carefully selected such that their sizes are very close. As shown in the scanning electron microscope (SEM) images in Figure S1, the particle size of Device \#1 is 907 nm $\times$ 837 nm, while the particle size of Device \#2 is 917 nm in diameter. The particle-sample separation can be controlled accurately using thin film deposition and atomic force microscopy (AFM) thickness calibration. Figure S2 shows the auto-oscillation results of Device \#1. The first localized mode reaches minimum FWHM linewidth of 20 MHz at RT, and 5 MHz at 80 K. The maximum power is 0.23 pW at RT, and 5.1 pW at 80 K. By comparing Figure S2 with Figure 2 (main text), the experimental results from two devices closely resemble each other with minimal device-to-device variation regarding mode structures and oscillator property of minimum linewidth. The spectral separations between localized modes are about the same. There are only two slight differences. First, the resonance frequencies of the localized modes in Device \#1 are 0.4 GHz smaller than the frequencies in Device \#2. This is due to the slightly larger moment of particle in Device \#1. Second, the bias currents in Device \#1 are 0.3 $\sim$ 0.5 mA higher than currents in Device \#2, and could be due to the slight device-to-device variations. Such closely resembling results support that the localized modes confined by particle-field are away from sample edges, and has avoided variations from the fabrication edge imperfections.

\begin{figure}[!htb]
	\includegraphics[width=0.7\columnwidth]{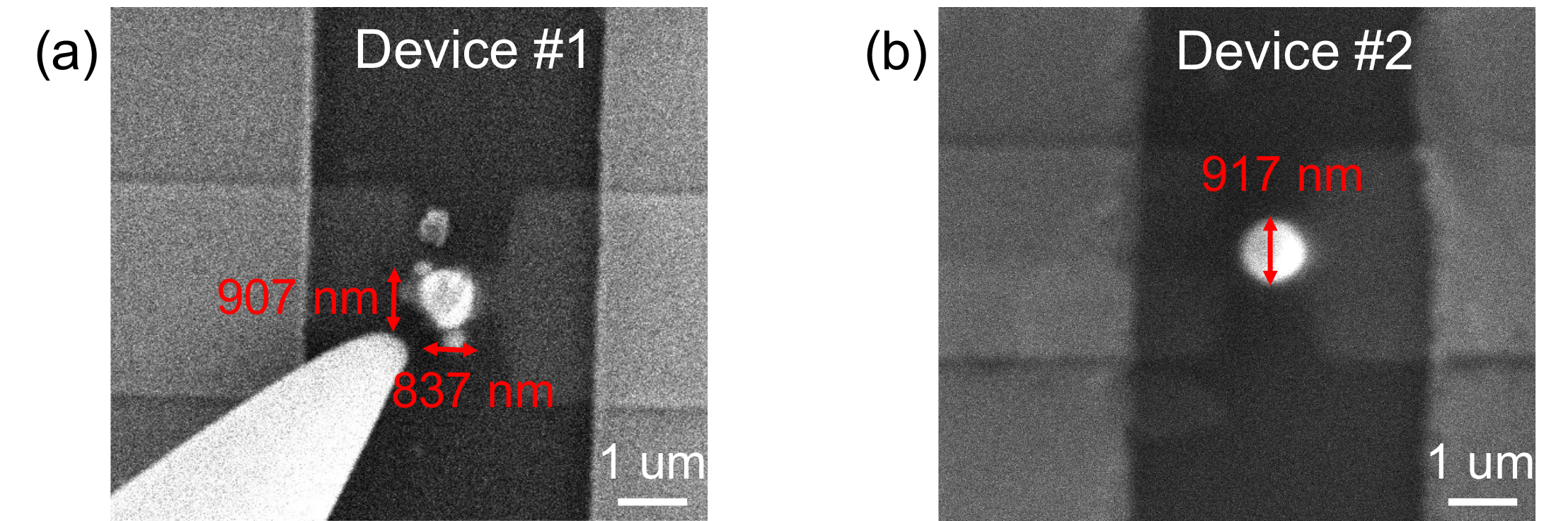}
	\caption{\label{FigureS1} Scanning electron micrograph of the two Py/Pt devices with Ni particles [Device \#1 and 2] used in the main text. The needle in the bottom left corner of (a) is a nano-manipulator in SEM.}
\end{figure}%

\begin{figure}[!htb]
	\includegraphics[width=0.8\columnwidth]{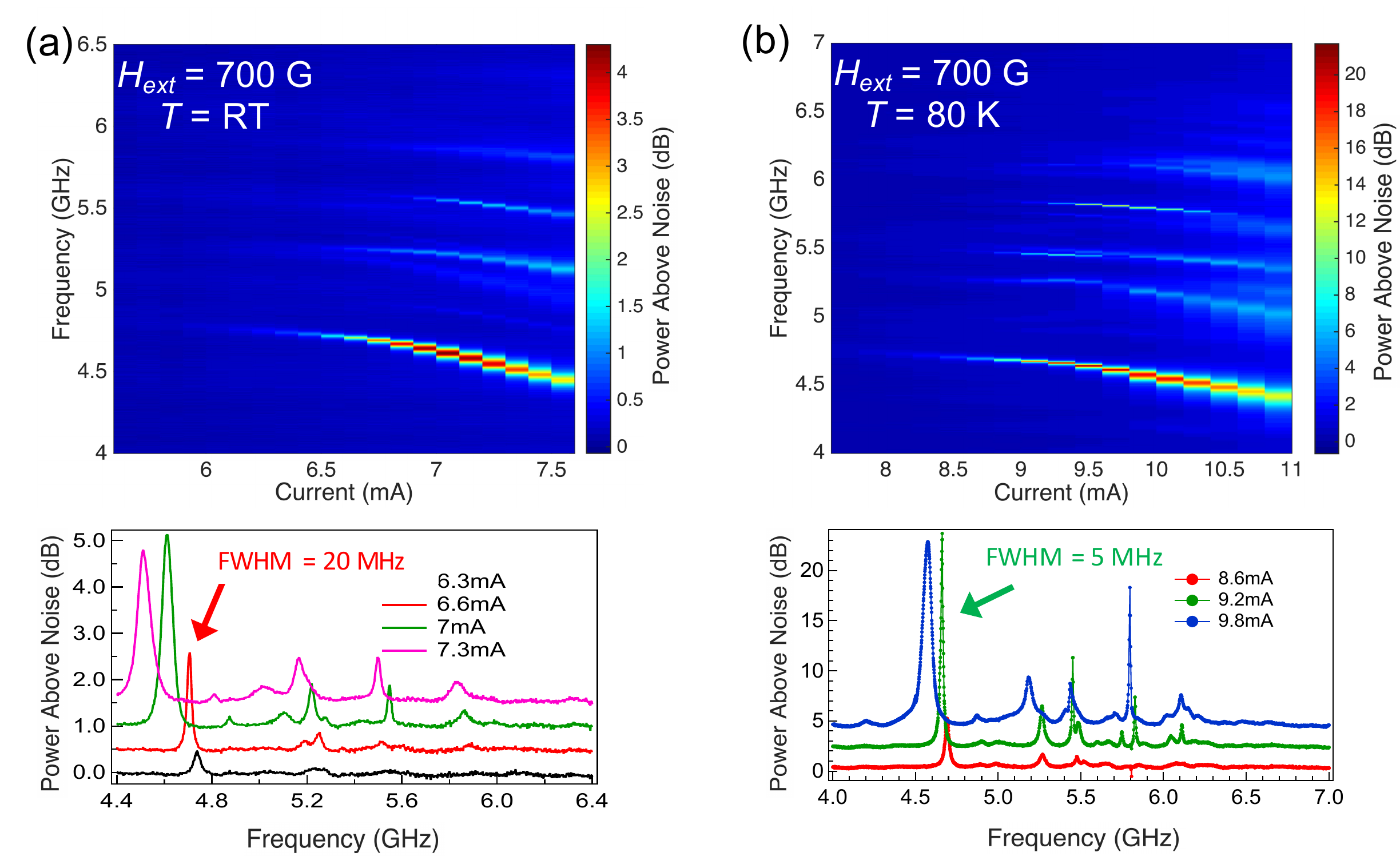}
	\caption{\label{FigureS2} Auto-oscillation color maps (as a function of bias current) and representative spectra of microwave emission generated by localized modes in Device \#1 at 700 G, (a) $T$ = RT and (b) $T$ = 80 K. The power baseline of the spectra are manually offset.}
\end{figure}%

\section*{B. Auto-oscillation color maps at different fields}
\begin{figure}[!htb]
	\includegraphics[width=0.8\columnwidth]{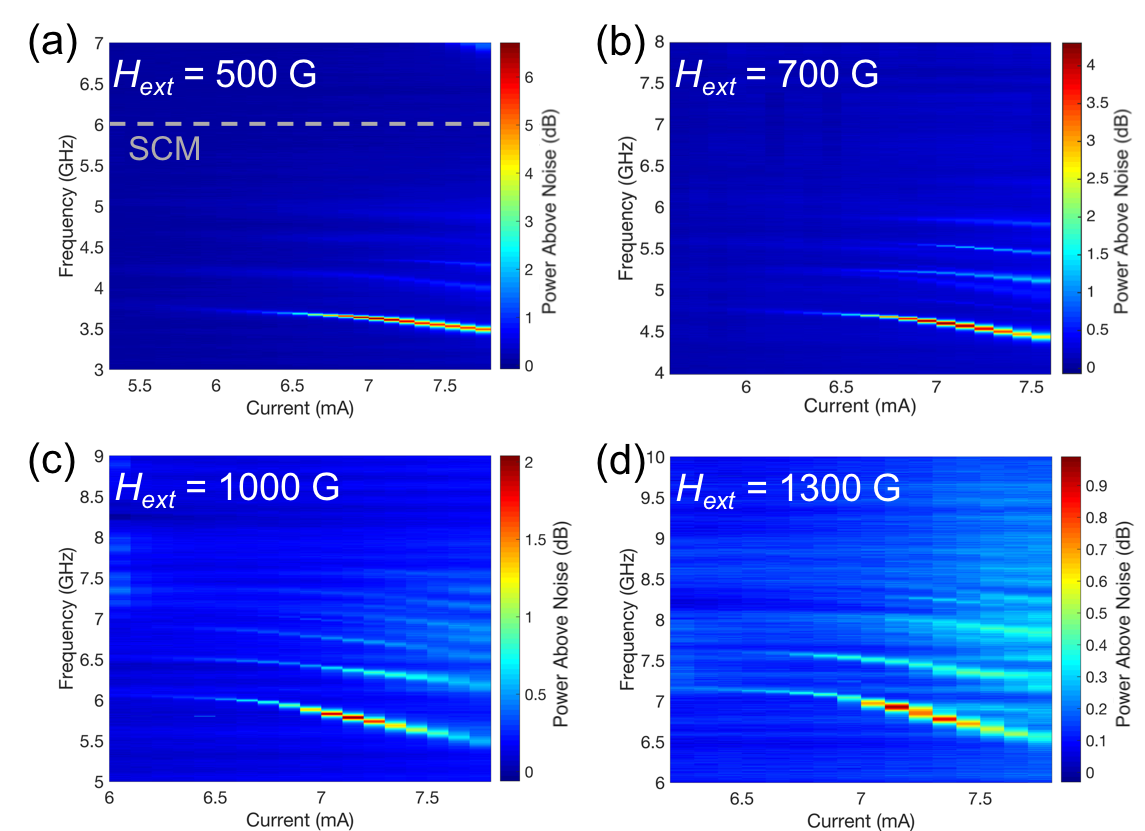}
	\caption{\label{FigureS3} Auto-oscillation color maps of microwave emission generated by localized modes in Device \#1 at different fields at RT. Dashed horizontal line (gray) in (a) indicates the frequency of sample-confined modes extracted from ST-FMR.}
\end{figure}%
\section*{C. Full spectrum and second harmonics}
Figure S4 shows the full spectrum with larger frequency range at 700 G, which gives information on which modes are excited in auto-oscillations. The group of the modes at frequency below 6.4 GHz are from localized modes. It is empty from 6.4 to 8.5 GHz, and then there are two high-frequency peaks. The frequencies of the two high-frequency peaks are twice the frequency of the two localized modes, denoted by markers, and therefore are second harmonics of those localized modes. The empty frequency region corresponds to the resonance conditions of the sample-confined modes (SCM). There are signals around those resonance conditions in ST-FMR spectra, but not in auto-oscillation spectra. This result can also be seen by comparing the ST-FMR result in Fig. 3(a) in main text and the auto-oscillation color map at 500 G in Fig. S3(a). Such results are reasonable because there are very high mode densities around those resonance conditions, with both sample-confined modes and higher-order localized modes. The degeneracies between a large number of modes and nonlinear four magnon scattering between multiple modes cause no modes into auto-oscillations.
\begin{figure}[!htb]
	\includegraphics[width=0.6\columnwidth]{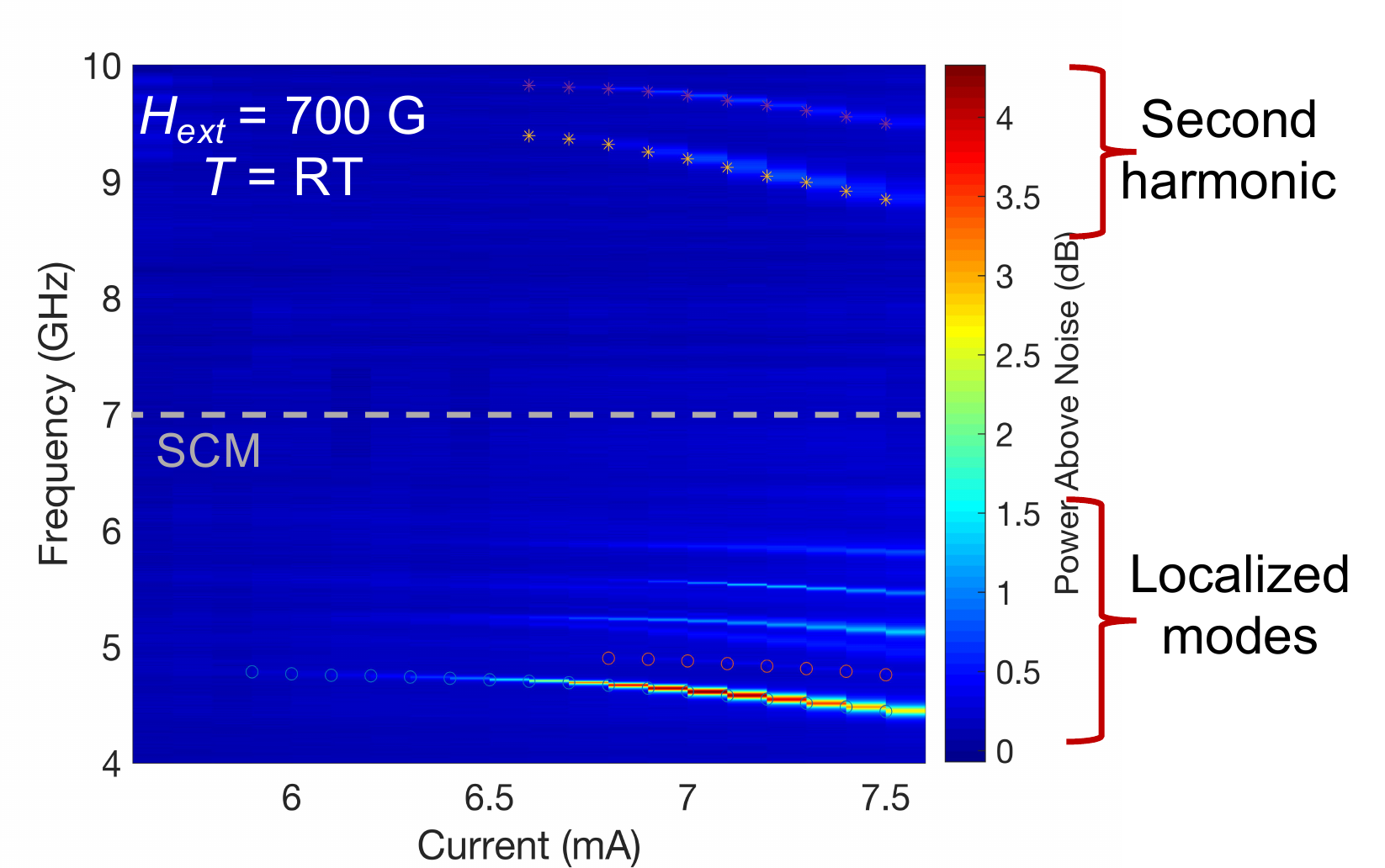}
	\caption{\label{FigureS4} Full spectrum with larger frequency range of the auto-oscillation color map of microwave emission generated by Device \#1 at 700 G. Dashed horizontal line (gray) indicates the frequency of sample-confined modes extracted from ST-FMR.}
\end{figure}%
\section*{D. Actual temperature $\textbf{T}_\textbf{a}$ vs. bath temperature $\textbf{T}_\textbf{b}$ \cite{liu_spectral_2013}}
\begin{figure}[!htb]
	\includegraphics[width=0.8\columnwidth]{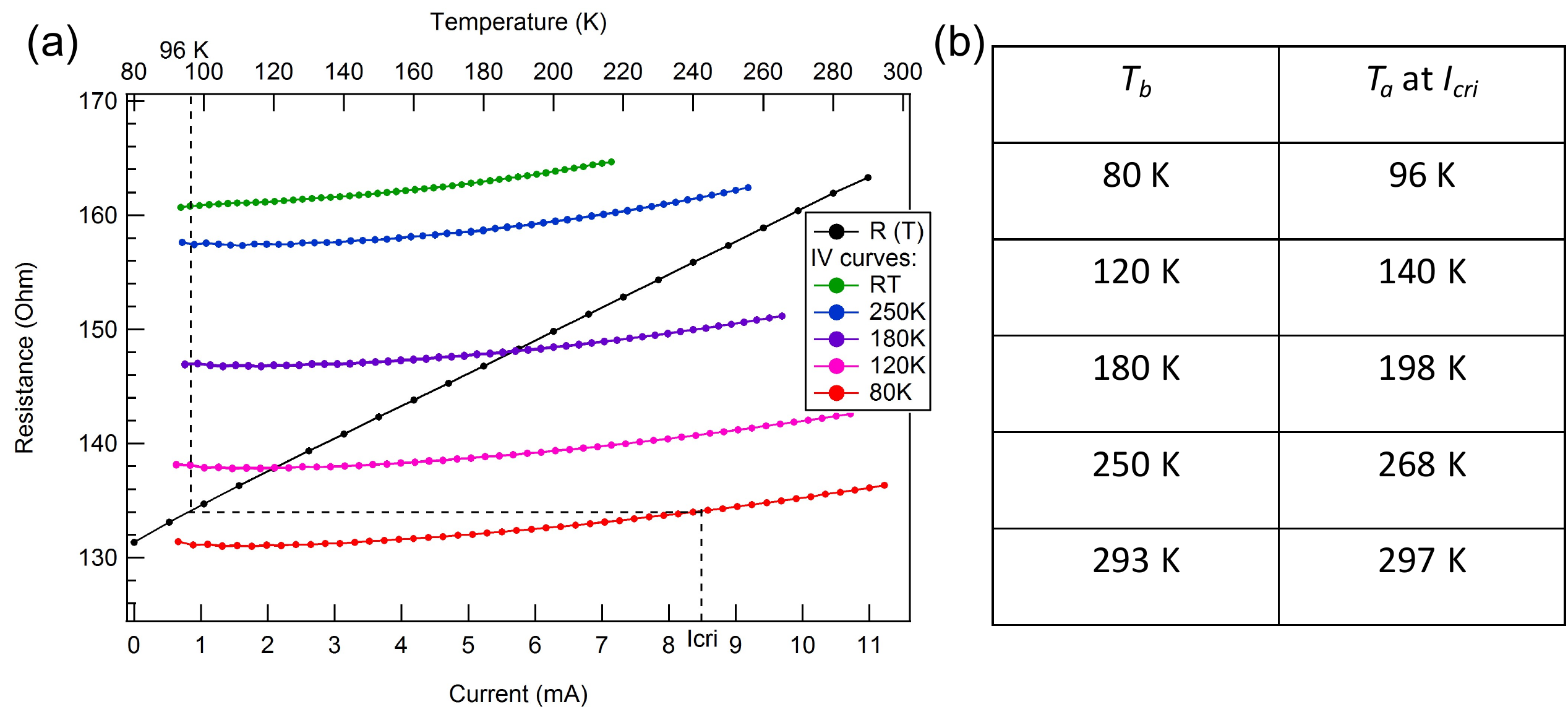}
	\caption{\label{FigureS5} (a) Resistance as a function of temperature and the resistance as a function of current at each bath temperature of Device \#2. (b) Actual temperature $T_a$ at critical current $I_{\text{cri}}$ for each bath temperature $T_b$, extracted from (a).}
\end{figure}%
\section*{E. Temperature-dependence color maps and representative spectra}
\begin{figure}[!htb]
	\includegraphics[width=0.93\columnwidth]{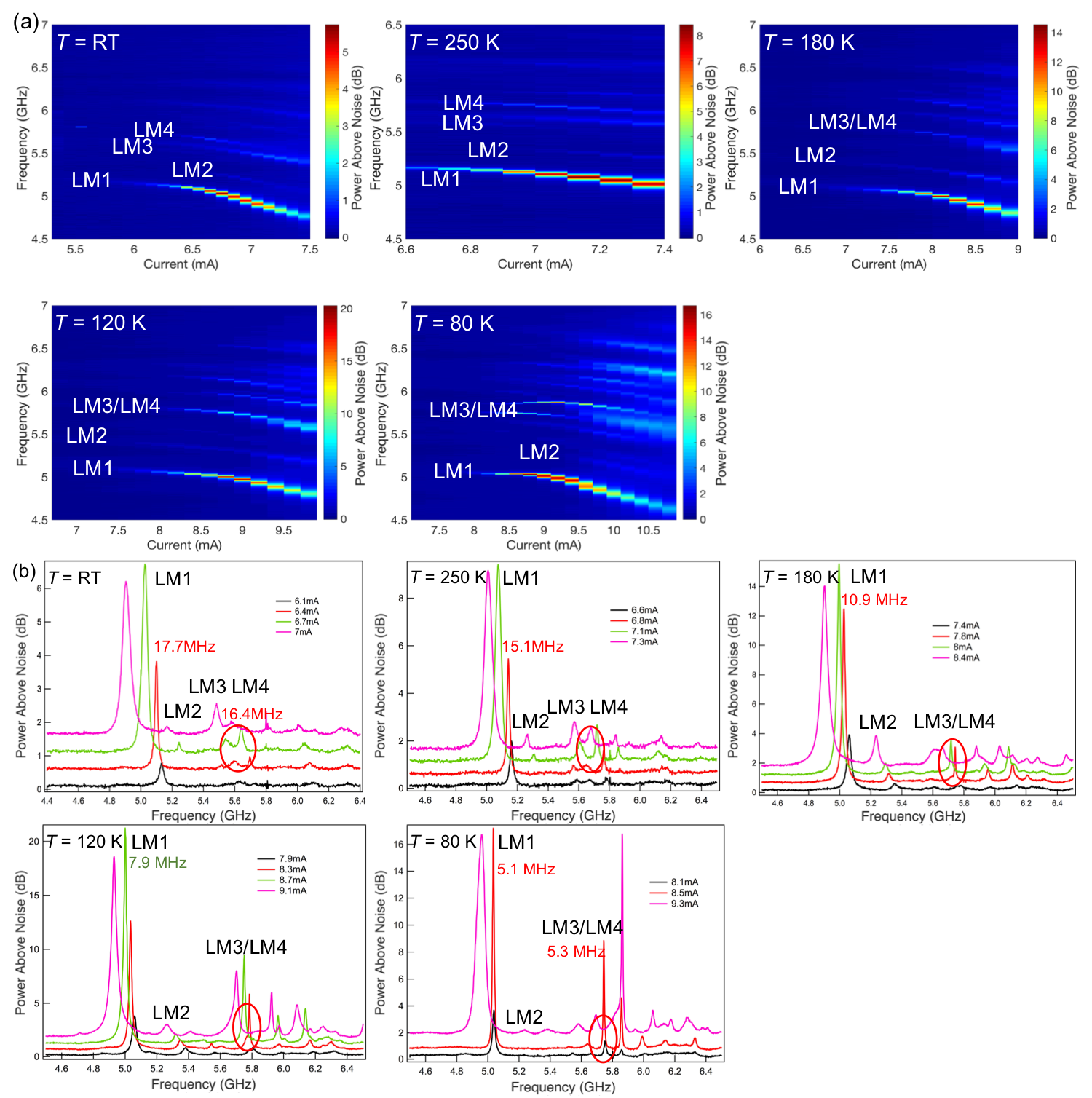}
	\caption{\label{FigureS6} Auto-oscillation (a) color maps and (b) representative spectra of microwave emission generated by localized modes in Device \#2 at different temperatures at 700 G. As seen in (b), a few higher-order localized modes grow into large signals, and are as sharp as the first localized mode. LM1 to LM4 are labeled on color maps and representative spectra, identified by comparing mode separations with micromagnetic simulation. The red circles locate LM3 and LM4 modes at RT and 250 K, and locate one of the LM3 or LM4 modes at 80 to 180 K. The minimum FWHM linewidth of localized modes at each temperature are labeled on graph.}
\end{figure}%
\section*{F. Bare strip data}
We measured bare 700 nm $\times$ 1 $\mu\rm m$ Py/Pt stripe without particle as supplementary measurements. The modes in a sub-micron stripe are bulk modes including quasi-uniform mode (QM), backward volume width modes at higher-field side of QM \cite{OscillatorDuan2015}, and surface modes at lower-field side of QM. Figure S7(a) shows ST-FMR spectra with increasing currents at 80 K, at $f$ = 6 GHz. We observed one main peak from quasi-uniform mode, and several smaller peaks from high-$k$ bulk modes. By comparing these spectra with spectra of Device \#1 with a particle in Figure 3(a) (main text), it is clear that qualitatively different modes are excited from those different spectra. The Mode A and Mode B peaks become very sharp at high currents, and will be onset for auto-oscillation first, as shown in auto-oscillation color maps at $H_{\text{ext}}$ = 580 G in Figure S7(b) and 700 G in Figure S7(c). Modes A, B, C and D are labeled in both ST-FMR and auto-oscillation spectra, correspondingly.
\begin{figure}[!htb]
	\includegraphics[width=1.0\columnwidth]{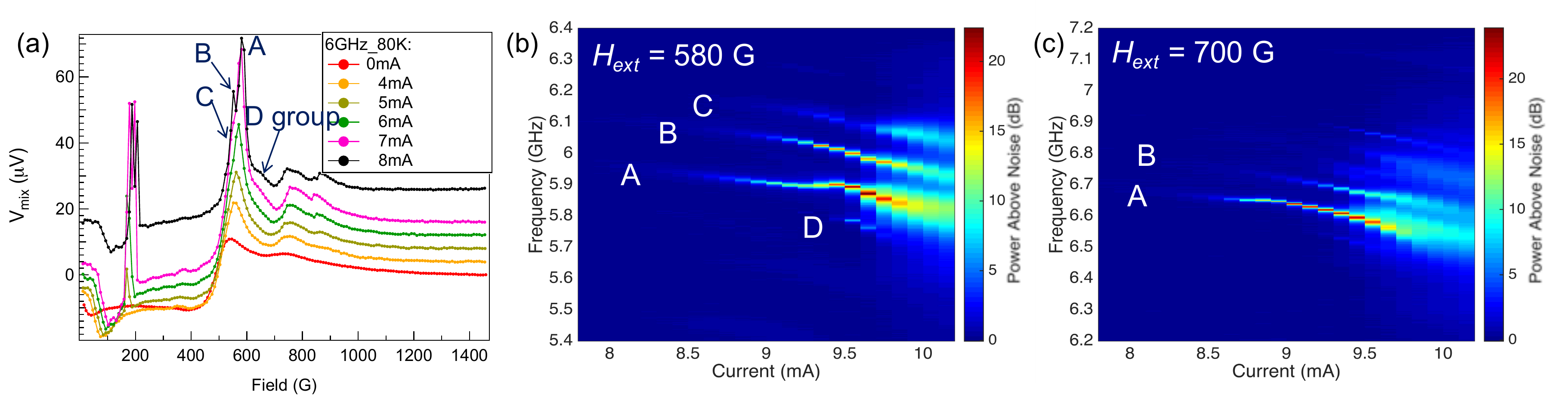}
	\caption{\label{FigureS7} Bare 700 nm $\times$ 1 $\mu\rm m$ Py/Pt stripe sample without particle (a) ST-FMR spectra for increasing currents, at 80 K, and $f$ = 6 GHz. Auto-oscillation color maps at 80 K, at (b) $H_{\text{ext}}$ = 580 G and (c) $H_{\text{ext}}$ = 700 G.}
\end{figure}%
\section*{G. Further details of Nickel particle gluing}
To glue the Ni particle to the center of the sample, we use optical microscope with micro-manipulators (Alessi REL-3200). A sharp glass tip, made by using a laser-based glass puller (Sutter Instrument P-2000) is inserted in the micro-manipulator. We put a drop of G1 epoxy, 1um-diameter Ni spherical particles (SkySpring Nanomaterials, Inc.) and our sample on a glass slide. We dip glass tip a few microns in to the epoxy to pick up a small droplet and put it on the sample. We then find a desired magnetic particle under the microscope and use another glass tip to pick it up with electrostatic forces. After putting the particle to the center of the sample, the exact position can be further adjusted by the glass tip from different directions. The device is then stored in $\text{N}_2$ box to avoid oxidization before the epoxy is cured. This gluing method is modified from a procedure from Ref. \cite{herman_studying_2011}.

In the measurements, the soft Ni particle aligns with the external field, with its dipole field anti-align with the external field, creating a spatial field well.

\section*{References}
%


\end{document}